\documentclass{article}

\usepackage{graphicx}  % standard LaTeX graphics tool;
\usepackage{amsmath}   % For getting proper math eqns
\usepackage{amssymb}   % Used for getting various symbols eg. gtrsim, lesssim etc.

\def\eq#1{{Eq.~(\ref{#1})}}

\newcommand{\bm}[1]{\boldsymbol{#1}}

\def\eq#1{{Eq.~(\ref{#1})}}

\newcommand{\LL}{Lanczos-Lovelock}

\newcommand{\mdof}{microscopic degrees of freedom}

\def\mbt{\bm}

  \title{Surface Density of  Spacetime Degrees of Freedom from Equipartition Law in  theories of Gravity}
  \author{T. Padmanabhan\\
  IUCAA, Pune University Campus,\\
  Ganeshkhind, Pune - 411 007.\\
  {\small {email: nabhan@iucaa.ernet.in}}
  }

  \date{}  %% This command  will supress printing the date.   If date is required, comment out this line.

\begin{document}
  
  \maketitle
  
\begin{abstract}
I show that the principle of equipartition, applied to area elements of  a surface $\partial\mathcal{V}$ which are in equilibrium at the local Davies-Unruh temperature, allows one to determine the surface number density of the microscopic spacetime  degrees of freedom in any diffeomorphism invariant theory of gravity. 
The entropy associated with these degrees of freedom matches with the Wald entropy for the theory.
 This result also allows one to attribute an entropy density to the spacetime in a natural manner. The field equations of the theory can then be  obtained by extremising this entropy. Moreover, when the microscopic degrees of freedom are in local thermal equilibrium, the spacetime entropy of a bulk region resides on its boundary.

\end{abstract}
 %%  Start the sections
 \section{Motivation and Summary: Equipartition of the `atoms of spacetime'}
 
 Considerable amount of theoretical evidence has accumulated over  years suggesting that gravity is better described as an emergent phenomenon like elasticity or fluid mechanics. (For a recent review, see \cite{rop}. This approach has a long history starting from the work of Sakharov; for a small sample of papers,  implementing and discussing this paradigm in different ways, see ref. \cite{others}.) Notable among these pieces of  evidence are the following facts: 
 \begin{itemize}
 \item 
 The  field equations of gravity reduce to a thermodynamic identity on the horizons in a wide variety of models much more general than just Einstein's gravity \cite{sphcqg, ronggen}. As pointed out first in \cite{surfaceaction}, and confirmed by several pieces of later work, \textit{the thermodynamic paradigm seems to be applicable to a wide class of theories much more general than Einstein gravity in 4-dimensions.}
 \item
 There are peculiar holographic relations between the surface and bulk terms in the action functionals describing several theories of gravity and the surface term in the action is closely related to entropy of horizons \cite{holo} in all these theories.
 \item
 It is possible to obtain the field equations of gravity --- again for a wide class of theories --- from purely thermodynamic considerations.
 \end{itemize}
  
  If this paradigm is correct, then
  the current situation regarding the dynamics of spacetime
  is similar to the state of affairs in our understanding of bulk matter before  Boltzmann.
  Spacetime consists of some microscopic degrees of freedom (``atoms of spacetime'') the dynamics of which will be governed by --- as yet unknown --- laws of quantum gravity. 
  In the long wavelength limit, we describe the dynamics of spacetime  by a set of variables (like metric, curvature etc.) the  evolution of which is governed by the field equations of gravity. This is similar to the description of, say, the flow of gas in terms of variables like fluid density,  velocity  etc. (which have no meaning at the microscopic scale) and the field equations of gravity are similar to laws of thermodynamics used in the description of bulk matter. As we said before, there is considerable amount of evidence to suggest that this is indeed the case.
  
  One key relation which connects the microscopic degrees of freedom with macroscopic thermodynamic variables is the law of equipartition which --- in the simplest context of a gas with constant temperature --- takes the form $E = (1/2) nk_B T$. In this relation,
   the energy $E$ and temperature $T$ are standard thermodynamic variables. But the variable $n$ giving the large, but finite, number of \mdof\ directly links the macroscopic thermodynamic description to the existence of \mdof.  In fact, $n$ has no meaning in a fully continuum theory with no discrete internal degrees of freedom. 
   
   In the thermodynamic paradigm of gravity, we assume that the spacetime has \mdof\ (just as Boltzmann assumed for a macroscopic body) and that the field equations of gravity in the continuum limit are to be obtained as the coarse-grained, thermodynamic, limit of the unknown microscopic laws. If this paradigm is correct, then there \textit{should} exist a relation akin to   $E = (1/2) nk_B T$ connecting the energy of a spacetime, temperature and the number of \mdof\ in a region of spacetime when some condition similar to that of thermodynamic equilibrium is satisfied. What is more, extensive thermodynamic potentials like for e.g. entropy $S$ of spacetime \mdof\ should scale as $S\propto n$ (The connection between this result and $S=\ln \Omega$ is described in the footnote 4 on page 11).
   
   Remarkably enough, \textit{such a relation does exist in any diffeomorphism invariant theory of gravity }\cite{cqgpap, equi} and the key purpose of this paper is to elaborate on this aspect. 
   
   In Sec. \ref{sec:eqlawmdof}, I will derive the equipartition law from the field equations of the theory and thus identify the number density of \mdof. I shall show that the relation $S\propto n$, with the $n$ obtained from the equipartition law, correctly reproduces the Wald entropy \cite{wald} for the horizons in any diffeomorphism invariant theory. This part of the paper \textit{assumes} the field equations and derives the equipartition law as well as the expression for spacetime entropy as a consequence. 
   In Sec.~\ref{sec:entdens},
   I will reverse the logic and indicate how the field equations of any diffeomorphism invariant  theory of gravity can be obtained from an entropy extremising principle. This has already been done in earlier papers \cite{aseemtp} and hence my discussion will be limited to stressing some new features and making the logical connections. Sec.\ref{speculate} comments on the possible manner in which the key results of the earlier sections can arise in a microscopic theory of gravity and the last section gives brief conclusions.
   
   Rest of the current section elaborates on these conceptual aspects and the reader who is familiar with them can skip ahead to Sec.  \ref{sec:eqlawmdof} after glancing at material around \eq{equione} and 
   \eq{equitwo}.

 \subsection{Boltzmann: ``If you can heat it, it has microstructure"}
 
 In the study of a macroscopic system like e.g., a gas  one uses two separate categories of physical variables. Those of the first category, like the bulk velocity field $\mathbf{v}(t,\mathbf{x})$ of the flowing gas, are direct generalizations of the corresponding mechanistic variables (viz., the velocity) used in the description of, say, a point particle moving under the influence of a potential in classical mechanics. Another example of such a variable is the pressure exerted by the gas on the walls of the container that is related to the force, which is also a familiar variable used in   point particle mechanics. 
 
 There is, however, a second category of variables which becomes necessary in the macroscopic description but are not used in the point particle mechanics. For example, we knew from prehistoric times that \textit{ bulk matter can be heated}.
 The  description of hot material bodies requires the introduction of concepts like temperature and entropy. Their behaviour was governed by the laws of thermodynamics, the microscopic origin of which was unclear for a long time.  These variables in classical thermodynamics (temperature, heat, entropy etc.) therefore acquire a special status different from, say, velocity field or pressure. They are useful and arise in the macroscopic limit,  when we study sufficiently large number of microscopic degrees of freedom collectively, but cannot be defined (usefully) for the microscopic point mechanics of particles.

In fact, not only that these variables (like entropy, temperature etc.) are useful, they are \textit{absolutely essential} for the thermodynamic description. Right from the early days, one knew, for example, that heat could be transfered between bodies and the laws of thermodynamics --- incorporating the flow of heat --- can be used to describe the behaviour of, e.g., heat engines.  
When  a hot gaseous body is coupled to, say, a movable piston in a heat engine, one can directly convert heat into mechanical energy. To describe such processes we will use
the standard, mechanistic, variables of the moving piston along with the thermodynamical variables describing the gas. 
So, it appeared that one needed two kinds of variables, purely mechanical ones (like velocity, pressure etc which are generalizations of variables used in point mechanics) and thermodynamic ones (like entropy, temperature, heat content etc. which did not have analogues in the context of point mechanics) for the proper description of macroscopic bodies.

It took the genius of Boltzmann to unify these two categories of variables into one. He realized that if one \textit{postulates} that material bodies have a large number of microscopic degrees of freedom, they can store up the energy which we call heat. That is, heat can  be interpreted as a form of motion of the microscopic degrees of freedom and entropy can be given an interpretation in terms of the unobserved (`coarse-grained') degrees of freedom. This allows one to reduce entire thermodynamic phenomena to mechanical phenomena of the microscopic degrees of freedom, thereby eliminating the need for two separate categories of variables.

A key relation that connects the microscopic and macroscopic descriptions is the law of equipartition of energy, say, in a gas. If the number of \mdof\ in certain amount of gas at temperature $T$ is $\Delta n$, then we expect:
\begin{equation}
E=\dfrac{1}{2}k_B \int  dn T
\label{equione}
\end{equation} 
 (The integral relation allows for the \mdof\ at different parts of gas to have different temperature.) This new variable $\Delta n$ appearing in \eq{equione}  has no significance in the absence of microscopic degrees of freedom since it specifically counts these degrees of freedom. Both $E$ and $T$ were standard variables used in thermodynamics but the introduction of $\Delta n$ to obtain the relation in \eq{equione} provides a direct link between the \mdof\ and the  macroscopic variables. All extensive thermodynamical variables contributed by the \mdof\ $\Delta n$ in certain amount of gas will now scale in proportion with $\Delta n$; in particular the entropy of a part of gas will satisfy $\Delta S\propto \Delta n$. We will see later that the equipartition law plays a key role in our description of spacetime.

In principle, once we relate thermodynamics  to the mechanics of \mdof, the laws of thermodynamics can be  derived by taking the suitable limit of the laws of microscopic physics. 
In practice, however, this is not an easy  task except for extremply simple systems. 
Fortunately, this is often unnecessary. What is essential, in practical contexts,  is  the expression for a suitable thermodynamic potential, say entropy ($S$) or free energy ($F$), in terms of appropriate macroscopic variables. By expressing the differentials $dS$ or $dF$ in terms of appropriate dynamical variables, (or  by extremising the thermodynamic potential), one can obtain the macroscopic laws governing the system. The thermodynamic description of matter in terms of these variables  turns out to be fairly universal and the differences between any two systems (say, helium and hydrogen gas) could always be incorporated in terms of a few well-chosen numbers (like e.g. specific heats or elastic constants). A law like $TdS=dE+PdV$, for example,  needed no outward modification either due to relativity or quantum theory, unlike other laws of physics. The effects of relativity or quatum theory only modifies the form of $S(E,V)$ thereby leading to appropriate description.  In this sense, thermodynamic description is remarkably devoid of specifics  and --- at the same time --- robust!

The key to this progress lies in the
Boltzmann's insight in postulating the microstructure for matter with a large but finite number of \mdof. 
Instead of treating heat, temperature etc as separate macroscopic variables, Boltzmann related them to the microscopic degrees of freedom for which --- at that time --- we did not have any direct observational evidence. Boltzmann \textit{inferred} the existence of microscopic degrees of freedom from the elementary fact that one can heat up matter! A hot system need to store the extra energy suppied to it internally and this \textit{demands} the existence of microscopic degrees of freedom. 

In short, Boltzmann said: ``\textit{If you can heat it, it has microstructure.}''
 
\subsection{And you can heat up the spacetime} 
 
\textit{Another physical system that can be heated is the spacetime.} To see this in an operational context, consider an observer who is at rest in the spacetime around a a spherical body of mass $M$. She can arrange matters such that the region around her has zero temperature as measured by a thermometer. Let the spherical body collapse and form a Schwarschild black hole. The thermometer will now indicate a non-zero temperature showing that the spacetime has been heated up. In fact, we do not even need a collapsing spherical body. If the observer accelerates through the inertial vacuum\footnote{Ideally, the acceleration has to be constant at all times; practically, the result holds approximately if the acceleration is constant for timescales sufficiently larger than $1/\kappa$; see e.g. \cite{dawoodvarg}} with a proper acceleration $\kappa$, the thermometer carried by the observer will indicate \cite{daviesunruh} a temperature $T=\kappa/2\pi$. Both these thought experiments show that spacetimes can be heated up just like, say, a body of gas. 

We stress that the temperature observed by, say, the accelerating thermometer is as real as that observered by an inertial thermometer inserted in a hot gas. The heat energy involved is also real and one can indeed heat up water using a black hole or by accelerating through the inertial vacuum. The universality of the phenomenon shows that it is the \textit{spacetime} which is hot; other material systems (like quantum fields) which reside in the spacetime reach thermal equilibrium with the hot spacetime and behave as systems at nonzero temperature.

Given the fact that spacetime --- just like a material body --- can be heated up, we assume, following Boltzmann, that spacetime has certain number density of  microscopic degrees of freedom. The exact nature of these degrees of freedom and their microscopic description are at present unknown, just as the exact nature
of atoms and molecules and their microscopic description were unknown when Boltzmann inferred their existence. This should not matter in the case of spacetime --- just as it did not matter in the case of bulk matter --- if  a suitable coarse-grained thermodynamic limit exists, which is robust and independent of the microscopic details. 

There is an acid test to this paradigm. Static spacetimes with  horizons can  be attributed both temperature and energy. Once we accept the existence of microscopic degrees of freedom, 
it seems reasonable to assume that they will reach equipartition at the given temperature.
 (The static nature of spacetime ensuring some kind of notion of equilibrium).
 Hence we expect a relation similar to \eq{equione} to hold in the static spacetimes; that is, we expect the gravitational energy $E$ to be expressible in the form
\begin{equation}
E=\frac{1}{2}k_B\int d n T
\label{equitwo}
\end{equation} 
where $\Delta n$ denotes the \mdof\ of the spacetime which are at temperature $T$. More importantly, such a relation will allow us to read off the density of \mdof\  of the spacetime
thereby providing a direct window to microphysics. We will see in Sec.~\ref{sec:eqlawmdof} that \textit{such an equipartition law arises in any diffeomorphism invariant theory of gravity whenever the field equations hold.} 

[This equipartition law was first obtained for Einstein's theory in ref.\cite{cqgpap} in the form $S=E/2T$. It was expressed in the explicit form of \eq{equitwo}  and elaborated upon in ref. \cite{equi} (see equation(12) of ref. \cite{equi}). For a sample of later work, exploring related ideas, see ref.\cite{equidisc}.]

The hot spacetimes also have a natural notion of entropy. 
In both the cases (collapsing body or accelerating observer), the region of spacetime accessible to the observer is limited by a causal horizon.  The existence of the horizon, blocking some information suggests that we could attribute  an entropy to the horizon. In the case of black hole horizon, it is generally accepted that the horizon \textit{is} endowed with an entropy, though the explicit expression for the entropy depends on the theory for gravity one adopts. (In Einstein's general relativity, the black hole entropy is proportional to the horizon area but not in more general theories). There is no general consensus as to whether other horizons, like the Rindler horizon or deSitter horizon, should be endowed with entropy but it seems inevitable from the thermodynamic point of view that all horizons have entropy. 

Again, if our ideas are correct, one should be able to determine the entropy of the static spacetime horizons by counting the microscopic degrees of freedom and using 
the extensivity in the form
$\Delta S\propto \Delta n$ where the $\Delta n$ can be read off from the equipartition law.
 (The connection between this result and $S=\ln \Omega$ is described in the footnote 4 on page 11.)
 The fact that it should match with the entropy obtained using other   prescriptions --- like first law of thermodynamics --- will act as a crucial consistency check of the approach. \textit{We will see that our approach indeed satisfies this consistency condition. }

Once we relate the entropy to the \mdof\ one would like to obtain the dynamics of the system from extremising the entropy. 
We assume that in the thermodynamic limit the spacetime can be described by a smooth differential manifold with some dimension $D$, suitable signature $(- + + + ....)$ and a metric. (This is far less ambitious a plan than trying to obtain spacetime itself as an emergent entity etc. I do not believe we understand the situation well enough to harbor such pretensions!). From the principle of equivalence, we infer that the gravitational field can be described by a metric tensor $g_{ab}$ so that the determination of spacetime dynamics reduces to obtaining suitable equations that determine this effective, `long-wavelength' variable. The existence of microstructure implies that the dynamics could be obtained by extremising a suitable thermodynamic potential, say, entropy $S[q_A]$ expressed in terms of appropriate variables $q_A$. 

There is one crucial ingredient we need to incorporate into the description at this stage: While we will see that there is a strong parallel between the normal thermodynamics  of a macroscopic body and the thermodynamics of the spacetime,   there is one  new feature which arises in the latter that needs to be emphasized. 
It is  clear from the  thought experiments introduced earlier (accelerating through the vacuum, collapsing a body to form a blackhole ....) that the  phenomena arising from heating up the spacetime are observer-dependent because horizons are observer dependent. An inertial observer (or an observer falling freely into a black hole) will attribute different thermal properties to the spacetime compared to an accelerated observer or an observer at rest outside the black hole \cite{comment2}.

As a result, different observers will attribute different thermodynamic properties to the same spacetime endowed with a particular geometrical structure (metric, curvature etc.). For example, a freely falling observer and a stationary observer will attribute different thermodynamic variables to a black hole spacetime. The variables $q_A$ we choose to express the entropy $S[q_A]$, say, should be capable of incorporating this observer dependence in a proper manner.
In particular, there is no need for these degrees of freedom to be same as the metric; but the extremum principle $\delta S[q_A]/\delta q_A=0 $ must constrain the \textit{background metric} under suitable circumstances. In fact, these degrees of freedom  $q_A$ itself could be different in different contexts, just as we can express the entropy in terms of different variables in normal thermodynamics, depending on our convenience.
 For example, if we are only interested in the study of static spacetimes with horizons, the entropy perceived by static observers will be the natural quantity to study. On the other hand, in a general spacetime, one has no  class of preferred observers. Then it turns out to be useful to work with local Rindler observers around any event and study the entropy functional in terms of the variables appropriate to them \cite{reviews2}. In such a case no Rindler observer is special and we need to maximize the entropy with respect to \textit{all} Rindler observers simultaneously. We will see that, such an approach emerges quite naturally from the study of equipartition and one can indeed obtain the field equations --- which, of course, are observer-independent --- from an entropy extremisation principle. We shall now see how these objectives can be achieved.

\section{Equipartition law and the density of \mdof\ }\label{sec:eqlawmdof}

The simplest context in which one can determine the density of \mdof\ is in the case of a static spacetime in Einstein's general relativity. To fix the ideas, we will begin with this case. 

Consider a a 3-volume $\mathcal{V}$  with a boundary $\partial \mathcal{V}$ in a static spacetime\footnote{We use the signature - + + +; Greek letters go over the spatial coordinates while Latin letters go over the spacetime coordinates. Except when otherwise indicated, we use units with $\hbar=c=k_B=1$ and $G=L_P^2=1$.} with  metric components $g_{00}=-N^2, g_{0\alpha}=0, g_{\alpha\beta}=h_{\alpha\beta}$.  An observer at rest in this spacetime  with four-velocity $u^a=\delta^a_0/N$ will have an acceleration $a^j=(0,a^\mu)$ where $a_\mu=(\partial_\mu N/N)$. In a static spacetime, it is easy to show \cite{cqgpap} that  
\begin{equation}
R_{ab}u^au^b=\nabla_ia^i=\frac{1}{N}D_\mu(Na^\mu)
\label{einsteinrab}
\end{equation}
where $D_\mu$ is the covariant derivative operator corresponding to the 3-space metric $h_{\alpha\beta}$.  
Einstein's field equations  relate the divergence of the acceleration to the source:
\begin{equation}
D_\mu(Na^\mu)\equiv 8\pi N\bar{T}_{ab}u^au^b
\equiv 4\pi \rho_{\rm Komar}
\label{arhokomar}
\end{equation}
where $\bar T_{ab}\equiv(T_{ab}-(1/2)g_{ab}T)$ and $\rho_{\rm Komar}$ is the (so called) Komar mass-energy density which is the source.
Integrating both sides of \eq{arhokomar} over  $\mathcal{V}$ and using the Gauss theorem gives:
\begin{equation}
E\equiv \int_{\mathcal{V}} d^3x\sqrt{h}\rho_{\rm Komar}
=\frac{1}{2}
 \int_{\partial\cal V}\frac{\sqrt{\sigma}\, d^2x}{L_P^2}\left\{\frac{N a^\mu n_\mu}{2\pi}
\right\}
\label{komarintegral}
\end{equation} 
where $\sigma$ is the determinant of the induced metric on $\partial \mathcal{V}$ and $n_\mu$ is the spatial normal to $\partial \mathcal{V}$. (We have temporarily restored $G=L_P^2\neq 1$ keeping $\hbar=c=k_B=1$.) We will now show, along the lines of \cite{equi}, that \textit{this result has a remarkable interpretation.}

To see this, choose $\partial \mathcal{V}$ to be a $N=$ constant surface so that the normal $n_\mu$ is in the direction of the acceleration and $a^\mu n_\mu= |\mbt{a}|$
is the magnitude of the acceleration. We can then introduce an effective Davies-Unruh temperature\cite{daviesunruh} $T =NT_{\rm loc}=(N a^\mu n_\mu /2\pi)=(N |\mbt{a}| /2\pi)$ in which the factor $N$  takes care of the Tolman redshift condition on the temperature \cite{comment1}. Hence \eq{komarintegral} is exactly in the form of the equipartition law in \eq{equione} (with $k_B=1$) where
\begin{equation}
E=\frac{1}{2}
 \int_{\partial\cal V}dn T;\quad \Delta n \equiv  \frac{\sqrt{\sigma}\,d^2 x}{ L_P^2}
 \label{idn}
\end{equation} 
That is, demanding the validity of (a) Einstein's equations and (b) law of equipartition we can \textit{determine} the number of \mdof\ in an element of area $\Delta A = \sqrt{\sigma} d^2 x$ on $\partial \mathcal{V}$ to be $ \sqrt{\sigma} d^2 x/ L_P^2$.  
If these microscopic degrees of freedom are in equilibrium at the temperature $T$, then the total equipartition energy contributed by all these degrees of freedom  on $\partial \mathcal{V}$ is equal to the total energy contained in the bulk volume enclosed by the surface --- which could be thought of as a realization of holographic principle.  
 One can, of course, rescale $\Delta n\to \Delta n/f$
 replacing  the factor $(1/2)$ in \eq{idn} by $(f/2)$ with some numerical factor $f$. We will stick with $f=1$ for simplicity; our results do not depend on this choice 
 \footnote{Incidentally, \eq{arhokomar} can be rewritten in a suggestive form as $D_\mu a^\mu=
 4\pi(\rho +\rho_g)$ where $\rho\equiv2\bar T_{ab}u^au^b$ and $\rho_g\equiv -(a^2/4\pi)$ is the energy density of gravitational field. Note that $\rho_g=-\pi T_{loc}^2$ acts as the source of gravity. The occurrence of $T_{loc}^2$ in the energy density rather than $T_{loc}^4$ has important implications for the nature of \mdof. This aspect is under investigation.}.

Remarkably enough, such an equipartition law exists for \textit{any} diffeomorphism invariant theory of gravity and allows us to identify the corresponding surface density $\Delta n/\sqrt{\sigma}d^2x$ of microscopic states in a given theory. Consider  a theory  based on the  action
 \begin{equation}
A=\int \sqrt{-g}\ d^Dx [L(R^a_{\phantom{a}bcd}, g^{ab})+L_{matt}(g^{ab},\phi_A)]
\end{equation} 
in $D$ dimensions where $L_{matt}(g^{ab},\phi_A)$ is a suitable matter Lagrangian involving some matter degrees of freedom $\phi_A$. Varying $g^{ab}$ in this action, with suitable boundary conditions,  leads to the field equations (see e.g. sec 3.5 of ref. \cite{rop} or, for a textbook description, chapter 15 of ref.\cite{gravitation}):
 \begin{equation}
\mathcal{G}_{ab}=P_a^{\phantom{a} cde} R_{bcde} - 2 \nabla^c \nabla^d P_{acdb} - \frac{1}{2} L g_{ab}
\equiv \mathcal{R}_{ab}-\frac{1}{2} L g_{ab}=\frac{1}{2}T_{ab}
\label{genEab}
\end{equation}
where
\begin{equation}
P^{abcd} \equiv \frac{\partial L}{\partial R_{abcd}}
\end{equation} 
The notation in terms of calligraphic fonts is suggested by the fact that, in Einstein's theory $\mathcal{R}_{ab}=R_{ab}/16\pi$ and $\mathcal{G}_{ab}=G_{ab}/16\pi$ with standard normalization.
The tensor $P^{abcd}$ has the same algebraic symmetries of the curvature tensor and hence can be indicated as $P^{ab}_{cd}$ without any ambiguity.
Any such theory of gravity, which is invariant under the diffeomorphism $x^a\to x^a+q^a$,  has a conserved Noether current
$J^a$ which depends on the vector field $q^a$ (see e.g. p.394, chapter 8 of \cite{gravitation} for a textbook description). This current can be explicitly computed  to be:
\begin{equation}
J^a = - 2 \nabla_b \left (P^{adbc} + P^{acbd} \right ) \nabla_c q_d + 2 P^{abcd} \nabla_b \nabla_c q_d - 4 q_d \nabla_b \nabla_c P^{abcd} 
\label{noedef1}
\end{equation} 
The conservation law $\nabla_a J^a=0$ implies that one can express the Noether current in terms of an antisymmetric Noether potential $J^{ab}$ with $J^a \equiv \nabla_b J^{ab}$.
The explicit expression for $J^{ab}$  which we will use is:
\begin{equation}
J^{ab} = 2 P^{abcd} \nabla_c q_d - 4 q_d \left(\nabla_c P^{abcd}\right)
\label{noedef}
\end{equation} 

In static spacetimes, we have a Killing vector $\xi^a$ corresponding to time translation invariance. If we take $q^a=\xi^a$, the expression for the Noether current is remarkably simple and we get $J^a=2\mathcal{R}^a_b\xi^b$. Hereafter, we shall denote by $J^a,J^{ab}$ etc., the Nother current and potential for this choice. Using the relations $J^a \equiv \nabla_b J^{ab}, \xi^a=Nu^a$ and the antisymmetry of $J^{ab}$ one can easily show that:
\begin{equation}
 2 \mathcal{R}_{ab} u^a u^b = \nabla_a (J^{ba} u_b N^{-1})
 \label{ruutotdiv}
\end{equation}  
 Next using the fact that, in a static spacetime $\nabla_i Q^i = N^{-1} D_\alpha (N Q^\alpha) $ for any static vector (with $\partial_0Q^i=0$) we can also write this relation as:
\begin{equation}
2N\mathcal{R}_{ab} u^a u^b  = D_\alpha (J^{b\alpha} u_b)
\end{equation} 
which is the generalization of \eq{einsteinrab} to an arbitrary theory of gravity. 
The integral version of this relation for a region $\mathcal{V}$ bounded by $\partial\mathcal{V}$ is given by
\begin{equation}
 \int_\mathcal{V} 2 N\mathcal{R}_{ab} u^a u^b \sqrt{h}\, d^{D-1}x 
= \int_{\partial\mathcal{V}}d^{D-2}x  \sqrt{\sigma}(n_iu_bJ^{bi}) 
= \int_{\partial\mathcal{V}}d^{D-2}x  \sqrt{\sigma} (N n_\alpha J^{\alpha 0})
\label{identity}
\end{equation}
where we have used $u_a=-N\delta_a^0$ and $J^{0\alpha}=-J^{\alpha 0}$. (The middle relation shows that the result is essentially an integral over $\partial\mathcal{V}$ of $J^{bi}d\sigma_{ib}$, where $d\sigma_{ib}=(1/2)n_{[i}u_{b]} \sqrt{\sigma}d^{D-2}x$.)
Further, the source for gravity in a general theory (analogous to Komar mass density) is defined through $\rho\equiv 4N\mathcal{R}_{ab} u^a u^b$.
On integrating $\rho$  over a region bounded by a $N=$ constant surface and using
\eq{identity}
 we get
\begin{equation}
E \equiv \int_\mathcal{V} \sqrt{h} d^{D-1}x \ \rho =2  \int_{\partial\mathcal{V}}d^{D-2}x  \sqrt{\sigma} (N n_\alpha J^{\alpha0})
\label{genequi}
\end{equation} 
This is the analogue of the equipartition law in \eq{komarintegral}. In Einstein's theory, $\mathcal{R}_{ab}=R_{ab}/16\pi$ and $J_{ab}=(16\pi)^{-1}\partial_{[a}\xi_{b]}$ giving $J^{\mu0}=a^\mu/8\pi$ which will reduce \eq{genequi} to \eq{komarintegral}. In a general theory, the expression for $\Delta n$ is not just proportional to the area and  $J^{\alpha0}$ encodes this difference. Our first aim is to understand this difference.

To do this, we note that the expression in \eq{genequi} simplifies to an interesting form in two contexts. 
Let us recall that the field equations of the theory, given by \eq{genEab} will contain higher than second order derivatives of the metric due to the $\nabla^c \nabla^d P_{acdb}$ term. This undesirable feature can be avoided by restricting  to the class of theories for which $\nabla_a P^{abcd}=0$, which are essentially the \LL\ models \cite{LL}.
 (Because of the symmetries of $P^{abcd}$, this condition implies that it is 
 divergence-free in all the indices.)
 In that case, we see from \eq{noedef} that $J^{ab}=2P^{abcd}\nabla_c\xi_d$
giving $J^{\alpha0}=4|\mbt{a}|P^{0\alpha}_{0\beta}n^\beta$ giving
\begin{equation}
 E=\int_{\partial\mathcal{V}}d^{D-2}x  \sqrt{\sigma} (16\pi P^{0\alpha}_{0\beta}n^\beta n_\alpha)
 \left(\frac{N|\mbt{a}|}{2\pi}\right)\equiv\frac{1}{2}\int dn T
 \label{equipartlaw}
\end{equation} 
where $T=N|\mbt{a}|/2\pi$ 
is the Davies-Unruh temperature
as before 
 (which depends only on the metric and \textit{not} on the field equations of the theory)
 but the number of microscopic degrees of freedom $\Delta n$ associated with an area element $\sqrt{\sigma}d^{D-2}x$ is now given by:
\begin{equation}
 \Delta n=32\pi P^{0\alpha}_{0\beta}n^\beta n_\alpha \sqrt{\sigma}d^{D-2}x
 =32\pi P^{ab}_{cd}\epsilon_{ab}\epsilon^{cd}\sqrt{\sigma}d^{D-2}x
\end{equation} 
where $\epsilon_{ab}\equiv (1/2)(u_an_b-u_bn_a)$ is the binormal to $\partial\mathcal{V}$. 

The second context in which the above analysis remains valid is when one deals with a \textit{general} theory but evaluates the surface integral in \eq{genequi} on the bifurcation horizon. In this case, we can again use the expression $J^{ab}=2P^{abcd}\nabla_c\xi_d$ because the additional term in \eq{noedef} vanishes on the bifurcation horizon on which $\xi^a=0$. We again get the same equipartition law in \eq{equipartlaw} on the horizon of any diffeomorphism invariant theory of gravity.
In either context, \eq{equipartlaw}  shows that the surface density of microscopic degrees of freedom  in a diffeomorphism invariant theory of gravity is given by
\begin{equation}
 \frac{dn}{\sqrt{\sigma}d^{D-2}x}=32\pi P^{ab}_{cd}\epsilon_{ab}\epsilon^{cd}
 \label{diffeoeqn}
\end{equation} 
when the field equations are satisfied. This is the key result of this paper.

The integrals over $\partial\mathcal{V}$ which occur in \eq{genequi} also arise in another context. In the action functional for gravity there will be a  surface term $A_{sur}$ arising from the spacetime integral of a total divergence in the Lagrangian $L_{sur}$. In the static case, it can be shown that (see eq.(172) of \cite{rop}), this surface term is given by
the integral of 
\begin{equation}
\sqrt{-g}L_{sur}=-\sqrt{h}(N\mathcal{R}_{ab}u^au^b)=-\frac{1}{2}\sqrt{h}D_\alpha(u_iJ^{i\alpha})
\end{equation}  
The integral of this term over spatial coordinates will lead to the surface term in \eq{genequi}. It is known that \cite{surfaceaction} one can obtain the field equations of the theory from a particular kind of variations of the surface term alone in the action and the above relation between $A_{sur}$ and the Noether potential connects these two concepts. This will be discussed in detail in a future publication.

\section{Entropy density of spacetime}\label{sec:entdens}

The density of \mdof\ obtained in \eq{diffeoeqn} suggests that the entropy associated with a \textit{general surface} in \LL\ models (which includes Einstein's theory) or the entropy associated with a horizon in a general theory  will be proportional to an integral over $P^{ab}_{cd}\epsilon_{ab}\epsilon^{cd}$. That is,\footnote{Note that in Einstein's theory, we get $\Delta n=\Delta A/L_P^2$. One usually considers this as arising due to dividing the area $\Delta A$ into $\Delta n$ patches of area $L_P^2$. If we attribute $f$ internal states to each patch, then the total number of microstates $\Delta\Omega$ will be $\Delta\Omega=f^{(\Delta n)}$ and $\Delta S=\ln \Delta\Omega\propto \Delta n$ which is how the extensivity $\Delta S\propto \Delta n$ arises.
In a more general theory, we replace $\Delta n=\Delta A/L_P^2$ by the expression in 
\eq{diffeoeqn}.}
\begin{equation}
S\propto\int_{\partial\mathcal{V}} dn \propto\int_{\partial\mathcal{V}}32\pi P^{ab}_{cd}\epsilon_{ab}\epsilon^{cd}\sqrt{\sigma}d^{D-2}x
\label{waldentro}
\end{equation} 
This is precisely the expression for Wald entropy \cite{wald} \textit{but we have obtained it using only the equipartition law}! We shall now examine the implications of this result in detail.

For a macroscopic system like a gas, one can obtain the dynamical equations from maximizing an entropy functional $S[q^A]$ expressed in terms of appropriate variables $q^A$. Analogously, one should be able to derive the field equations of gravity  from maximizing a suitable  entropy functional of spacetime \cite{aseemtp}.
 However, as we said before, there is one crucial difference between thermodynamics of gases and thermodynamics of spacetime. We know that the same spacetime can exhibit different thermal behaviour to different observers and hence one would expect the entropy functional etc. to take  different forms in different contexts. We will need to apply the maximization principle to a \textit{class} of observers to obtain the dynamical equations.  We will now describe how one can obtain a suitable form of entropy functional.

To motivate this, consider a static spacetime with a  
bifurcation horizon $\mathcal{H}$ given by the surface $N^2 \equiv - \xi^a \xi_a =0$. The horizon  temperature $T \equiv\beta^{-1}= \kappa/2\pi$ where $\kappa$ is the surface gravity.
Since the Wald entropy \cite{wald} of the horizon is essentially the Noether charge (multiplied by $\beta$), 
we will attribute (along the lines of \cite{entdenspacetime}) the Noether charge \textit{density}  $\beta J_b u^b $ (multiplied by $\beta$) as the 
 entropy density of the spacetime as perceived by the static observers with four velocity $u^a= \xi^a/N$, so that the total entropy is  
\begin{equation}
S_{\rm grav}[u^i] = \beta\int_\mathcal{V}  J_b u^b \sqrt{h}\, d^{D-1}x 
\label{defs1}
\end{equation} 
Using $J^a = 2 \mathcal{R}^a_b \xi^b$ and \eq{identity}  and integrating the expression over a region bounded by the $N=$ constant surface, it is easy to see that
\begin{equation} 
S=\frac{1}{2}\, \beta E
\end{equation} 
which is a statement of equipartition, discussed in detail earlier in ref. \cite{cqgpap, equi}. Further, if we take $\partial\mathcal{V}$ to be the horizon $\mathcal{H}$ and use $\beta T=1$, we get the horizon entropy to be
\begin{equation}
 S=\frac{1}{4}\int_\mathcal{H} dn
 =\frac{1}{4}\int_{\mathcal{H}}32\pi P^{ab}_{cd}\epsilon_{ab}\epsilon^{cd}\sqrt{\sigma}d^{D-2}x
\end{equation} 
which is the standard expression for Wald entropy \cite{wald} in a general theory thereby justifying the choice in \eq{defs1}. This ansatz in \eq{defs1} also fixes the proportionality constant in \eq{waldentro} to be $1/4$. It is easy to see that we get one quarter of horizon area in Einstein's theory. 

We started by assuming the validity of field equations and then obtained the equipartition relation in \eq{genequi}. This allowed us to determine the number density of \mdof\ and thus a possible expression for entropy density. We can now close the logical loop by taking the form of the entropy functional as the starting point and obtaining the field equations. If the thermodynamic interpretation of gravity is correct, one should be able to obtain get the field equations from extremizing the spacetime entropy of a set of \mdof.

 To do this, we have to recast the expression for $S$ in \eq{defs1} as a \textit{spacetime} integral (rather than spatial integral) and generalize it to a context in which the spacetime has no special attribute (like static nature). The first task is easy.
 The  spacetime entropy in \eq{defs1} can \textit{also} be expressed in the form
 \begin{equation}
S_{\rm grav}[u^i] = \beta\int_\mathcal{V}  J_b u^b \sqrt{h}\, d^{D-1}x 
 = \int 2\mathcal{R}_{ab} u^a u^b  \sqrt{-g}\, d^{D}x
 \label{defs}
\end{equation}  
by restricting the time integration to the range $(0,\beta)$ which can be justified by using the Euclidean continuation of the static spacetime with a horizon in which the time coordinate is periodic with period $\beta$. But since the last expression in \eq{defs} does not involve a $\beta$, we are no longer confined to a spacetime with a horizon. Further, in a general non-static spacetime, we do not have any special class of observers who can be used to define the vector field $u^a$. 
So we need to generalize this notion and introduce a suitable vector field in its place. This issue is conceptually more involved  but --- fortunately --- has already been addressed \cite{aseemtp,reviews2}.
We know from the study of accelerated observers
 in flat spacetime 
 that the null surfaces $X=\pm T$ in the inertial coordinates $(T,\mathbf{X})$ appear as the local horizon to the observer moving on the uniformly accelerated trajectory $X^2-T^2=\kappa^{-2}$ and has to be endowed with temperature and entropy. 
 One can  introduce a local inertial frame around any event and local Rindler observers 
 moving along a hyperbolic trajectory in the locally inertial coordinate systems
  around any event.
  These local Rindler observers will perceive the null surfaces,  generated by the null vectors at that event, as local Rindler horizons endowed with thermal properties.
  This suggests that a suitable variable describing local patches of null surfaces can play the role of $q_A$ in the entropy functional $S[q_A]$. The simplest such choice will be the set of all null vectors on the spacetime with some (as yet undetermined) metric. Denoting  a generic member of this set by $k^a$ we can express the spacetime entropy as a functional $S[k^a]$ obtained by integrating the entropy density $s[k^a]$ over the spacetime volume. The \textit{total} entropy density will, of course, be the sum of $s[k^a]$ and the matter entropy density.
One can equivalently think of this procedure as attributing the entropy to the normal displacements of null surfaces.
(A more detailed justification for this approach can be given in terms of local Rindler observers; see e.g. ref. \cite{rop,reviews2}).

 This suggests that the expression for gravitational entropy of spacetime can be taken, in the general context,
to be \cite{entdenspacetime}
\begin{equation}
S_{\rm grav}[k^i] \propto \int \sqrt{-g} \, d^Dx\mathcal{R}_{ab} k^a k^b
\label{Sdefgen}
\end{equation}  
It can be shown \cite{rop,aseemtp} that maximizing $(S_{\rm grav} + S_{\rm matter} )$ for all null vectors $k^a$ simultaneously\footnote{As explained in detail in ref. \cite{reviews2}, this procedure is conceptually identical to using local inertial frames to determine the kinematics of gravity, viz. how gravity couples to matter. To do that, we choose local inertial frames around each event in spacetime and insist that the equations of motion (or the action functional) for matter should reduce to the special relativistic form in \textit{all} the local inertial frames. Similarly here, to determine the dynamics of gravity, we choose local Rindler frames around each event and insist that the entropy should be an extremum in \textit{all}  local Rindler frames. This is why we need to demand that the extremum principle is valid for all $k^a$.} leads to  
the  field equations in \eq{genEab}.
 That is, the field equations in any diffeomorphism invariant theory can be obtained from an entropy maximization principle. This provides a conceptual connection between equipartition and the spacetime entropy density.
 
If one accepts $\beta J^a=\beta (2\mathcal{R}^a_b\xi^b)$ as the  entropy current in the local Rindler frame, where $\xi^a$ is the approximate Killing vector producing translations in the rindler time coordinate, it is possible to obtain the field equations in several different ways. All these procedures (described in \cite{rop,reviews2,entdenspacetime}) use a suitable vector $v^a$ which satisfies $\xi^a v_a=0$ on the local Rindler horizon $\mathcal{H}$ and obtains the relation 
\begin{equation}
J^av_a= (2\mathcal{R}^a_b\xi^b v_a)=T^a_b\xi^b v_a\quad (\rm{on}\ \mathcal{H})
\label{keyidea}
\end{equation} 
from thermodynamic arguments. It is then possible to obtain the field equations, except for a cosmological constant, from this equation. 

For example, consider the  entropy flux $\delta S$ through a small patch $(\sqrt{\sigma}d^{D-2}x)$ on the stretched horizon at $N=\epsilon$ (where $\epsilon$ is an infinitesimal quantity), in a propertime interval $(Ndt)$. 
Since $\beta J^a$ is the entropy current, this flux
will be given by 
$\delta S=\beta J^an_a(\sqrt{\sigma}d^{D-2}x)(Ndt)$ where $n^a$ is the spatial normal to the stretched horizon. On the other hand, the flux of energy through the same surface is 
$ \delta E=T^{ab}\xi_bn_a(\sqrt{\sigma}d^{D-2}x)(Ndt)$, leading to a heat transfer of
$\beta\delta E$. Equating it to the entropy flux, we immediately get \eq{keyidea} for the choice $v^a=n^a$. This approach --- described in \cite{entdenspacetime} --- which uses Noether \textit{current} and is local in construction,  shows that one can obtain the field equations from an entropy balance argument, if one is prepared to accept $\beta J^a$ as the entropy current --- which, of course, needs independent justification. (The equipartition result strengthens this idea but we needed to assume field equations to get equipartition law; the same holds for Wald entropy construction which is also an on-shell result.) 

One can also attempt \cite{mauliksarkar} to express the same result as a relation between integrated quantities  by, say,  associating the Wald entropy --- which is the integral of $(\beta/2)J^{ab}\epsilon_{ab}(\sqrt{\sigma}d^{D-2}x)$ (where $\epsilon_{ab}$ is the binormal to the stretched horizon) --- with two $(D-2)$ dimensional surfaces $N=\epsilon,t=t_1$ and  $N=\epsilon,t=t_2$ on the stretched horizon.  The difference in the entropy associated with the two surfaces is given by the flux of $J^a=\nabla_b J^{ab}$ through the boundaries  and one can try to relate it to matter energy flux through the stretched horizon. This approach has two problems: First, it also needs an independent justification  as to why the integral of $(\beta/2)J^{ab}\epsilon_{ab}(\sqrt{\sigma}d^{D-2}x)$ is entropy since the notion of Wald entropy assumes validity of field equations. Second,  since the notion of local Rindler frames is only valid over a finite region, any integral one uses has to be independently justified (or should be confined to sufficiently small region, which reduces it to the previous, local, approach). But since the local approach \textit{does} work, the nonlocal approach in terms of integrated quantities will also lead to the same result (viz. the derivation of field equations) if done correctly. In particular, the correct approach will \textit{not} have any extra dissipative terms.

The expression in \eq{Sdefgen} for $S_{\rm grav}$ in a general theory can be expressed in an alternate form by separating out a total divergence. Direct computation shows that 
\begin{equation}
2 \mathcal{R}_{ab} k^a k^b \equiv 4 \nabla_c\left[ P^{cd}_{ab} k^a \nabla_d k^b\right] + \mathcal{S}_{\rm grav}[k^i]
\end{equation} 
where
\begin{equation}
\mathcal{S}_{\rm grav}[k^i] = 4\left[ P^{cd}_{ab}\nabla_c k^a \nabla_d k^b + (k^b \nabla_c k^a)\nabla_d P^{cd}_{ab} +k_a k_c \nabla_b \nabla_d P^{abcd}\right]
\label{calsgrav}
\end{equation} 
is a quadratic expression in $k^a$ and its derivatives. One can obtain \cite{rop,sfwu} the same field equations in \eq{genEab} by using $\mathcal{S}_{\rm grav}$ as the entropy density instead of $2 \mathcal{R}_{ab} k^a k^b$. 
In the case of \LL\ models, the expression in \eq{calsgrav} simplifies considerably and we find that the gravitational entropy density of spacetime is a quadratic expression
$
\mathcal{S}_{\rm grav} \propto P^{cd}_{ab}\nabla_c k^a \nabla_d k^b 
$ which was investigated earlier in ref.\cite{aseemtp}.

There are two interesting features worth mentioning about the expression for entropy in \eq{calsgrav}. First, it can be shown that (see e.g., sec 7.4 of ref.\cite{rop})
when the field equations hold, the entropy $\mathcal{S}_{\rm grav}[k^i]$ of a volume $\mathcal{V}$ will reside on its surface $\partial\mathcal{V}$. Second, and probably more interesting fact is that
  in any static spacetime $\mathcal{S}_{\rm grav}[u^i]$ (defined using the four velocity of static observers instead of null vectors $k^i$) will be a total divergence identically --- that is, even off-shell. This follows from the fact that, in any static spacetime
   $\mathcal{R}_{ab}u^au^b$ itself is a total divergence (see \eq{ruutotdiv}). Using  this expression and simplifying the terms, one can show, after some algebra that
   \begin{equation}
\mathcal{S}_{\rm grav}[u^i] =-\nabla_b[4u_au_d\nabla_cP^{abcd}]
\end{equation} 
Hence, in any static geometry, the  entropy perceived by static observers also resides on the surface.  Both these results  generalizes the notion of holography beyond Einstein's theory.

\section{Speculations on the connection with the microscopic theory}
\label{speculate}

The discussion so far has been devoid of unnecessary speculations which was possible because we took a `top-down' approach. The key result, expressed, for example, in the form of \eq{diffeoeqn}, could be derived from classical theory with the only quantum mechanical input being the formula for Davies-Unruh temperature. In fact the $\hbar$ enters the expression only through the formula $k_bT=(\hbar/c)(a/2\pi)$. It nicely combines with $G$ in \textit{classical} gravitational field equations and leads to $G\hbar/c^3$, one of the central constants in \textit{quantum} gravity. So we did not make any unwarranted speculations or dubious assumptions from the domain of quantum gravity to obtain \eq{diffeoeqn}.

One may, however, be curious to know where a result like \eq{diffeoeqn} fit in the broader picture. For example, one might wonder how such a result might emerge from a microscopic candidate theory of quantum gravity like for example, string theory (for a texbook description see e.g.,\cite{book1}) or loop quantum gravity (for a texbook description  see e.g.,\cite{book2}). Given the fact that \textit{all} such candidate theories involve fair amount of leaps of faith, it is not possible to discuss this issue without entering into some speculation. We will do so in this section, just to indicate the realm of possibilities but it must be stressed that the rest of the paper is completely independent of the discussion in this section.

To begin with, it is clear that \textit{our result in \eq{diffeoeqn} should not depend on the details of the final, correct, theory of quantum gravity}. This is exactly analogous to the fact that the thermodynamics of gases, say, is fairly independent of the microscopic details of the  gaseous system one is considering. The microscopic description of Helium gas can be quite different from that of Argon but at the level of ideal gas equation or thermdynamic processes we can ignore most of these details  and quantify the relevant differences in terms of a few parameters like, e.g., specific heat. Similarly, any model of quantum gravity which has (i) correct classical limit and (ii) is consistent with 
Davies-Unruh acceleration temperature, will lead to our \eq{diffeoeqn}. Since one would expect any quantum gravity model to satisfy these two conditions (i) and (ii) above --- and because our results only depend on these two features --- our `top-down' approach will meet the `bottom-up' approach of such a theory in the overlap domain. So if string theory or loop-quantum-gravity or dynamical triangulations or any other candidate model for quantum gravity satisfies (i) and (ii), it will lead to our result. (On the other hand if the quantum gravity model does not satisfy the conditions (i) and (ii), such a candidate model is probably wrong in any case.).

Thus \eq{diffeoeqn} does not --- and it is not expected to, either --- put a serious constraint on the microscopic models. This is again in the same spirit as the fact that law of equipartition in statistical mechanics does not put any constraint on the atomic nature of matter. After all, one cannot derive Schrodinger equation for hydrogen atom, knowing the equipartition law for hydrogen gas.

Even though it seems reasonable to accept that thermodynamic limit contains far less information about the system than a microscopic description will provide, one could still wonder what could be the possible route or mechanism by which a \textit{wide class} of candidate models in quantum gravity (all of which satisfies (i) and (ii) in the last paragraph) can lead to a relation like \eq{diffeoeqn} in the appropriate limit. Obviously, such a mechanism should use sufficiently general theoretical concepts which are independent of the specific details of the microscopic models. This question can be answered by indicating one possibility by which such results can emerge but such a discussion is necessarily speculative. We will now briefly describe this possibility, based on the results of several earlier papers, especially ref. \cite{Squant}. (This is also motivated by the author's view that the currently available candidate models for quantum gravity are failures and a fresh outlook needs to be developed.)

Let us first consider Einstein's theory in which the equipartition law assigns $A/L_P^2$ degrees of freedom to a proper area element $A$ so that $A\approx n L_P^2$ for $n\gg 1$. (This corresponds to the semiclassical limit in which one can meaningfully talk about proper area elements etc. in terms of a coarse grained semiclassical metric.) This is, of course, nothing but area quantization in the asymptotic limit. Though LQG probably leads to such an exact result, in the asymptotic limit we are interested in, it can be obtained from the Bohr-Sommerfeld quantization condition applied to horizons. In fact, Bekenstein conjectured \cite{Bekenstein1} long back that, in a quantum theory, the black hole {\it area} would be represented by a quantum operator with a discrete spectrum of eigenvalues.
 (Bekenstein showed that the area of a classical black hole behaves like an adiabatic invariant, and so, according to Ehrenfest's theorem, the corresponding quantum operator must have a discrete spectrum.) Extending these ideas to local Rindler horizons treated as the limit of a stretched horizon, one can understand how a result like quantization of any spatial area element can arise in a microscopic theory.
 
 A more interesting situation emerges when we go beyond the Einstein gravity to \LL\ models.
  In Einstein's gravity, entropy of the horizon is proportional to its area. Hence one could equivalently claim that it is the gravitational \textit{entropy} which has an equidistant spectrum.
But, when one considers the natural generalization of Einstein gravity to \LL\ models the proportionality between the area and Wald entropy breaks down. (There is, of course, no generalization of LQG for \LL\ models.) The question then arises as to whether it is the quantum of area or quantum of entropy (if at all either) which arises in a natural manner  in these models. This question was addressed in ref. \cite{Squant} where it was shown that in the \LL\ models, it is indeed \textit{the entropy that is quantized} with an equidistant spectrum. This matches nicely with the fact that it is the quantity in the right hand side of \eq{diffeoeqn} --- which, as we pointed out in Sec.\ref{sec:entdens}, leads to the Wald entropy of horizons ---  that takes integral values in the equipartition law. Thus one can alternatively interpret our result as a form of entropy quantization. 

We can now indicate a sufficiently general  `mechanism' by which any microscopic model of quantum gravity can possibly lead to such a quantization condition, in terms of  two ingredients. First one is the peculiar `holographic' structure of the action functionals in \LL\ models \cite{holo,sanvedholo}. 
The action functionals in \textit{all} these theories can be separated into a bulk and surface term and the surface term has the structure of an integral over $d(pq)$. It can also be shown \cite{rop,holo} that the same `$d(pq)$' structure emerges for the on-shell action functional in all \LL\ models; that is,
\begin{equation}
A|_{\rm on-shell}\propto \int_{{V}}d^Dx \partial_i( \Pi^{ijk}g_{jk})=
\int_{\partial \mathcal{V}}d\Sigma_i\; \Pi^{ijk}g_{jk}
\label{onshellpdq}
\end{equation} 
where $\Pi^{ijk}$ is the suitably defined canonical momenta corresponding to $g_{jk}$.
The second ingredient is the fact that, in all these theories, the above expression for the action leads to the Wald entropy of the horizon. (For example, in Einstein gravity, the surface term in action, evaluated on a horizon will give one quarter of the area \cite{holo}.)

In the semiclassical limit, Bohr-Sommerfeld quantization condition requires the integral of $d(pq)$ should be equal to $2\pi n$. Since the action in \eq{onshellpdq} has this `$d(pq)$' structure, it follows that the Bohr-Sommerfeld condition reduces to  $A|_{\rm on-shell}=2\pi n$. Since $A|_{\rm on-shell}$  is also equal to Wald entropy, we get
\begin{equation}
S_{\mathrm{Wald}}=A|_{\rm on-shell}=2\pi n
\label{bohr}
\end{equation}
The Bohr-Sommerfeld quantization condition, of course, was the same used originally by Bekenstein and others to argue for the area quantization of the black hole horizon but in the more general context of \LL\ models it appears as entropy quantisation \cite{Squant}.
(It is also possible to argue that in the
 \textit{semiclassical} limit, the
 on-shell value of the action will be related to the phase of the semiclassical wave function
$
\Psi \propto \exp \left(iA|_{\rm on-shell}\right)
$. If the semiclassical wave function describing the quantum geometry, relevant for a local Rindler observer, is obtained by integrating out the degrees of freedom beyond the horizon inaccessible to the observer then one can argue that \cite{Squant}, this phase should be $2\pi n$ in the asymptotic limit leading to  $A|_{\rm on-shell}=2\pi n$. Given the conceptual ambiguities related to interpretation of `wave function of geometry', it is probably clearer to invoke Bohr-Sommerfeld condition.)
To summarize, the three facets of the theory: (i) the structure of the gravitational action
functional (ii) the equality of  on-shell gravitational action functional and Wald entropy
and (iii) the Bohr-Sommerfeld quantization condition, combine together to provide a possible backdrop in which \eq{diffeoeqn} can arise in any microscopic theory.

We once again stress that the above analysis is speculative and involves ill-understood concepts from semiclassical gravity but the rest of the paper is completely independent of these speculations. 
The purpose of this discussion is only to point out one possible, tentative, scenario in which
results like \eq{diffeoeqn} can arise from a microscopic theory of quantum gravity. In the absence of the latter, it is not possible to do better or provide a more concrete or unique discussion.

\section{Conclusions}

\begin{table}
\caption{Equipartition law applied to a macroscopic body and spacetime}
\begin{center}
% use packages: array
\begin{tabular}{lll}
\textbf{System} & \textbf{Macroscopic body} & \textbf{Spacetime} \\ 
\hline
\noalign{\bigskip}
Can the system be hot? & Yes & Yes \\ 
\noalign{\bigskip}
Can it transfer heat? & Yes; for e.g., hot gas  & Yes; water at rest  \\ 
& can be used to &in Rindler spacetime \\
&  heat up water&will get heated up\\
\noalign{\bigskip}
How could the heat  & The body must have   & Spacetime must have \\ 
energy be stored &microscopic degrees  &microscopic  degrees  \\
in the system? &of freedom&of freedom\\
\noalign{\bigskip}
How many microscopic & Equipartition law & Equipartition law  \\ 
 degrees of freedom  is& $dn=dE/(1/2)k_BT$&$dn=dE/(1/2)k_BT$\\
 required to store energy&&\\
 $dE$ at temperature $T$?&&\\
\noalign{\bigskip}
Can we read of $dn$? & Yes; when thermal  
 & Yes; when static field \\
&equilibrium holds; &  equations  hold; depends  \\ 
&depends on the body&on the theory of gravity \\
\noalign{\bigskip}
Expression for entropy & $\Delta S\propto \Delta n$ & $\Delta S\propto \Delta n$ \\ 
\noalign{\bigskip}
Does this entropy match& Yes & Yes \\
 with the expressions  &&\\
 obtained by other methods? &&\\
\noalign{\bigskip}
How does one close the  & Use the entropy   & Use the entropy   \\
loop on dynamics? &  extremisation to   & extremisation to  \\
& obtain thermodynamical  &obtain gravitational \\
&equations&field equations\\
\noalign{\bigskip}
Are the thermal phenomena  & Yes; however,  & No! \\
 observer-independent?& see \cite{comment2}.&
\end{tabular}
\end{center}
\end{table}

\bigskip

We see that the the equipartition law allows one to identify the number density of microscopic degrees of freedom on a constant redshift surface. Using this one can define the entropy of the horizon in a general theory of gravity (which agrees with Wald entropy). This definition of entropy, recast in terms of null vector fields,  allows us to obtain the field equation by an entropy extremisation principle. These facts further strengthen the idea that gravity is an emergent phenomenon and spacetime thermodynamics --- \textit{which extends far beyond Einstein gravity} --- is more fundamental. Rather than repeat the arguments given in the text, I have summarized them in the form of a table.

The thermodynamic approach to gravity brings to the centrestage the  Noether current and potential, which were not considered major players in the theory of gravity before. In the case of a continuum fluid or elastic solid,  the displacement $x^\alpha\to x^\alpha+q^\alpha(x)$ is considered as   an elastic deformation  of the solid and the physics can be formulated in terms of how the thermodynamic potentials (like the entropy) change under such displacement.  Similarly, in the case of spacetime, one could think of
$J^a$ and $J^{ab}$ as providing the response of the spacetime entropy density to the `deformation' of the spacetime $x^a\to x^a+q^a$. Then \eq{noedef1} and \eq{noedef}
give the response in terms of the dependence of $J^{ab}$ on $q^a$ and its gradients. For example, treating $q^a$ and $\nabla_bq_a$ as independent at any event one can write
\begin{equation}
P^{abcd}=\frac{1}{2}\frac{\partial J^{ab}}{\partial\nabla_cq_d}
\end{equation} 
which completely determines the structure of the theory. Since $J^{ab}d\sigma_{ab}$ gives the number density of microscopic degrees of freedom, the above relation connects $P^{abcd}$ to the response of \mdof\ on a surface to spacetime displacements. We can now formulate the 
theory in terms of this response function. For example, in the $m$-th order \LL\ model in $D-$ dimension the field equations are just
\begin{equation}
R_{jabc}\left(\frac{\partial J^{ia}}{\partial\nabla_bq_c}\right)=\bar T^i_j
\end{equation} 
with $\bar T^i_j\equiv [T^i_j -(1/(D-2m))T\delta^i_j]$. More details of this approach towards dynamics will be described in a future publication.

\section*{Acknowledgments}

I thank D.Kothawala and S.Kolekar for useful discussions and comments on the draft.


\begin{thebibliography}{100}

%%%%%%%%%%%%%%%%%%%%%%%%%%%%%%%%%%%%%%%%%%%%%%%%%%%%

\bibitem{rop} 
Padmanabhan T.,  2010, \textit{Thermodynamical Aspects of Gravity: New insights}, Rep. Prog. Phys., Rep. Prog. Phys. \textbf{73} (2010) 046901, [arXiv:0911.5004].

\bibitem{others}
Sakharov A D  1968  \textit{Sov. Phys. Dokl.} {\bf 12} 1040;
Jacobson T  1995 \textit{Phys. Rev. Lett.} \textbf{75}  1260;
Volovik G E  2001 \textit{Phys.Rept.}, \textbf{351} 195; 
Visser M 2002 \textit{Mod.Phys.Lett}. \textbf{A17}  977; 
Barcelo C et al.  2001 \textit{Int.J.Mod.Phys.} \textbf{D10}799;
Volovik G E  2003 \textit{The universe in a helium droplet}, (Oxford University Press); 
Jannes G 2009 \textit{Emergent gravity: the BEC paradigm} [arXiv:0907.2839];
Yang H S and Sivakumar M 2009 \textit{Emergent Gravity from Quantized Spacetime}                 [arXiv:0908.2809];
Makela J  \textit{Quantum-Mechanical Model of Spacetime} [arXiv:0805.3952];
Elizalde E and   Pedro J S 2008 \textit{Phys.Rev.} D \textbf{78} 061501;
Hogan C J 2008 \textit{Phys.Rev.} D \textbf{77} 104031;
Bamba K et al. 2009 \textit{Equivalence of modified gravity equation to the Clausius                   relation} [arXiv:0909.4397];
Chirco G and    Liberati  S 2009 \textit{Non-equilibrium Thermodynamics of Spacetime: 
      the  Role of Gravitational Dissipation} [arXiv:0909.4194];
Sindoni L Girelli F and  Liberati S 2009   \textit{Emergent gravitational dynamics
        in Bose-Einstein condensates}  [arXiv:0909.5391];
Bamba K  Geng C and   Tsujikawa S 2009  \textit{Equilibrium thermodynamics in 
        modified gravitational theories} [arXiv:0909.2159];
Barcelo C,  Liberati S and Visser M 2005
 \textit{Living Rev.Rel.} \textbf{8} No. 12 [gr-qc/0505065];
 F. Piazza, arXiv:0910.4677. 

\bibitem{sphcqg} 
T. Padmanabhan, \textit{Class.Quan.Grav.}\textbf{ 19}, 5387 (2002). [gr-qc/0204019]

\bibitem{ronggen}
Cai R. G. and Kim  S. P.  (2005), \textit{JHEP}, {\bf 0502}, 050  [hep-th/0501055];
A. Paranjape, S. Sarkar,  T. Padmanabhan,  \textit{Phys.Rev.}, \textbf{D 74}, 104015 (2006) [hep-th/0607240]; 
Akbar M. and Cai  R. G. (2006), \textit{Phys. Lett.},  B {\bf 635}, 7  [hep-th/0602156];
 D. Kothawala, S. Sarkar,  T. Padmanabhan,  \textit{Phys. Letts,} \textbf{B 652}, 338-342 (2007) [gr-qc/0701002]; 
Akbar M. and Cai  R. G.  (2007), \textit{Phys. Rev.},   {\bf D 75}, 084003  [hep-th/0609128];
Cai R. G. and Cao L. M. (2007),  \textit{Nucl. Phys.},  B {\bf 785}, 135 [hep-th/0612144];
Sheykhi A.,  Wang B. and Cai R. G.,  (2007), \textit{Nucl. Phys.},  B {\bf 779}, 1 [hep-th/0701198];
Sheykhi Wang A. B. and Cai R. G. (2007), \textit{Phys. Rev.}, D {\bf 76}, 023515  [hep-th/0701261];
Cai  R. G. (2008),  \textit{Prog. Theor. Phys. Suppl.}, {\bf 172}, 100 [arXiv:0712.2142];
Akbar M. and Cai  R. G. (2007), \textit{Phys. Lett.},  B {\bf 648}, 243   [gr-qc/0612089];
Cai R. G. and Cao L. M. (2007),  \textit{Phys. Rev.},  D {\bf 75}, 064008  [gr-qc/0611071];
Gong Y. and Wang A. (2007), \textit{Phys. Rev. Lett.}, {\bf 99}, 211301   [arXiv:0704.0793];
Wu S. F., Yang G. H. and Zhang P. M.  [arXiv:0710.5394];
Cai R. G., Cao L. M. and Hu Y. P. (2008), \textit{JHEP}, 0808:090 [arXiv:0807.1232]; 
Wu S. F., Wang B. and Yang G. H. (2008), \textit{Nucl. Phys.},  B {\bf 799}, 330 [arXiv:0711.1209];
Wu S. F. et al. (2008), \textit{Class. Quant. Grav.}, \textbf{25}, 235018 [arXiv:0801.2688];
Zhu T., Ren J. R. and Mo S. F.   [arXiv:0805.1162];
Cai R. G., Cao L. M. and Hu Y. P. (2009), \textit{Class. Quant. Grav.}, {\bf 26}, 155018 [arXiv:0809.1554]. 
 D.Kothawala,  T. Padmanabhan, \textit{ Phys. Rev.,}  \textbf{ D 79} , 104020 (2009) [arXiv:0904.0215] 
 
 \bibitem{surfaceaction}
T. Padmanabhan, \textit{Dark Energy: Mystery of the Millennium,}  Albert Einstein Century International Conference, Paris, 18-22 July 2005, AIP Conference Proceedings  861, Pages 858-866, [astro-ph/0603114].

 \bibitem{holo}
 T. Padmanabhan,  \textit{Mod.Phys.Letts}. \textbf{A 17}, 1147 (2002). [hep-th/0205278]; 
                   \textit{Gen.Rel.Grav.,}  \textbf{34}  2029-2035 (2002) [gr-qc/0205090];   
                 \textit{Brazilian Jour.Phys. (Special Issue) } \textbf{35}, 362 (2005) [gr-qc/0412068]; 
A. Mukhopadhyay,  T. Padmanabhan, \textit{Phys.Rev.,} \textbf{D 74}, 124023 (2006) [hep-th/0608120];
 F. Caravelli, L. Modesto, [arXiv:1001.4364].
 
 \bibitem{cqgpap}
 Padmanabhan  T., (2004),   \textit{Class.Quan.Grav.}, \textbf{21}, 4485  [gr-qc/0308070].
 

\bibitem{equi} 
Padmanabhan T.,(2010), \textit{Equipartition of energy in the horizon degrees of freedom and the emergence of gravity} 
Mod.Phys.Letts.A (in press),  [arXiv:0912.3165].

\bibitem{wald} 
Wald R. M.,  \textit{Phys. Rev. D}  (1993), {\bf  48}  3427, [gr-qc/9307038];
Iyer V.  and R. M. Wald, (1995), \textit{Phys. Rev. D} {\bf 52}  4430, [gr-qc/9503052].

\bibitem{aseemtp} 
Padmanabhan T.,   (2008), \textit{Gen.Rel.Grav.},   \textbf{40}, 529-564 [arXiv:0705.2533];
Padmanabhan T.  and Paranjape A., (2007), \textit{Phys.Rev.} D, \textbf{75}, 064004 [gr-qc/0701003].



\bibitem{daviesunruh}
Davies P C W 1975  \textit{J. Phys.} A \textbf{8}  609--616; Unruh W G  1976 \textit{Phys. Rev.} D \textbf{14}  870.


\bibitem{dawoodvarg}
D. Kothawala, T. Padmanabhan,  \textit{Response of Unruh-DeWitt detector with time-dependent acceleration}, [arXiv:0911.1017]


\bibitem{equidisc}
E. P. Verlinde, arXiv:1001.0785;
K. Ropotenko, arXiv:0911.5635;
Rong-Gen Cai, Li-Ming Cao, Nobuyoshi Ohta, \textit{Phys.Rev.} \textbf{D 81}, 061501 (2010) [arXiv:1001.3470];
Lee Smolin, arXiv:1001.3668;
Fu-Wen Shu, Yungui Gong, arXiv:1001.3237;
T. Padmanabhan, arXiv:1001.3380;
Jarmo Makea, arXiv:1001.3808;
Miao Li, Yi Wang, arXiv:1001.4466;
Changjun Gao,  arXiv:1001.4585;
Yi Zhang, Yun-gui Gong, Zong-Hong Zhu, arXiv:1001.4677;
Hristu Culetu, arXiv:1001.4740;
Yi Wang, arXiv:1001.4786;
Tower Wang, arXiv:1001.4965;
 Shao-Wen Wei, Yu-Xiao Liu, Yong-Qiang Wang, arXiv:1001.5238;  
  Yi Ling, Jian-Pin Wu, arXiv:1001.5324;
  Jae-Weon Lee, Hyeong-Chan Kim, Jungjai Lee, arXiv:1001.5445;arXiv:1002.4568;
   Liu Zhao, arXiv:1002.0488;
 Jerzy Kowalski-Glikman, arXiv:1002.1035;     
 Yu-Xiao Liu, Yong-Qiang Wang, Shao-Wen Wei,  arXiv:1002.1062;
 Rong-Gen Cai, Li-Ming Cao, Nobuyoshi Ohta, arXiv:1002.1136;
  Alessandro Pesci, arXiv:1002.1257;
 Yu Tian, Xiaoning Wu, arXiv:1002.1275;
 Yun Soo Myung, Yong-Wan Kim, arXiv:1002.2292;
 I.V. Vancea, M.A. Santos, arXiv:1002.2454;
R.A. Konoplya,arXiv:1002.2818;
Hristu Culetu, arXiv:1002.3876;
 Subir Ghosh, arXiv:1003.0285;
 Joakim Munkhammar, arXiv:1003.1262;
 Xiao-Gang He, Bo-Qiang Ma, arXiv:1003.1625;
Jae-Weon Lee, arXiv:1003.1878; arXiv:1003.4464 ;
Rabin Banerjee, Bibhas Ranjan Majhi, arXiv:1003.2312;
Xiao-Gang He, Bo-Qiang Ma, arXiv:1003.2510;
Yi-Fu Cai, Jie Liu, Hong Li, arXiv:1003.4526.
 \bibitem{reviews2}
Padmanabhan T., \textit{A Physical Interpretation of Gravitational Field Equations}, 
[arXiv:0911.1403];
Padmanabhan T., \textit{A Dialogue on the Nature of Gravity},	[arXiv:0910.0839v2]. 

\bibitem{comment2}
The situation becomes more complex --- and interesting --- if
an observer has a `normal' thermodynamical system, say, a box of gas at some temperature,
and accelerates through the inertial vacuum carrying it.
Then the thermal behaviour attributed to the gas by the observer will be a convolution of `normal' thermal behaviour and those arising from the acceleration temperature of spacetime, as perceived by the observer. Therefore even the `normal' thermodynamics now acquires \cite{reviews2} a new level of observer dependence in non-inertial frames. This is not often emphasised but a little thought shows that it is inevitable.



\bibitem{comment1}

In fact, the reason for choosing $\partial \mathcal{V}$ to be a $N=$ constant surface 
  is to ensure a constant redshift factor between the temperatures
 attributed to different area elements. Also note that $N|\mbt{a}|$ tends to a finite, constant, surface gravity $\kappa$ on a black hole horizon.

\bibitem{gravitation} T.Padmanabhan (2010) \textit{Gravitation: Foundations and Frontiers}, Cambridge University Press, UK.

 

\bibitem{LL} 
Lanczos C.  1932 {\it Z. Phys.} 
            {\bf 73} 147; 
Lanczos C. 1938 {\it Annals Math.} 
            {\bf 39} 842; 
Lovelock D.  1971 {\it J. Math. Phys.} 
            {\bf 12} 498. 

\bibitem{entdenspacetime}
 T. Padmanabhan, Int.Jour.Mod.Phys. \textbf{D18 }, 2189 (2009); [arXiv:0903.1254]
 
\bibitem{mauliksarkar}
M.K. Parikh, S. Sarkar, arXiv:0903.1176


\bibitem{sfwu} 
Shao-Feng Wu et.al,  (2010), \textit{Phys.Rev.D} \textbf{81}, 044010, [arXiv:0909.1367].

\bibitem{book1}  
J. Polchinski, \textit{String Theory, Vol. 1,2} (Cambridge University Press, 2005); 
B. Zwiebach, \textit{A First Course in String Theory} (Cambridge University Press; 2nd edition, 2009).

\bibitem{book2}
C. Rovelli, \textit{Quantum Gravity,} (Cambridge University Press, 2007);
T. Thiemann, \textit{Modern Canonical Quantum General Relativity,} (Cambridge University Press, 2007).



\bibitem{Squant}
 Dawood Kothawala,  T. Padmanabhan , Sudipta Sarkar,  Phys. Rev., \textbf{D78}, 104018 (2008) [arXiv:0807.1481]

%%%%%%%%%%%%%%%%%%%%%%%%%%%%%%%%%%%%%%%%%%%%%%%%%%%%%%%%%%%%%%%%%%%%%%%%%%%%%%%
\bibitem{Bekenstein1}
J.D. Bekenstein, \textit{Lett. Nuovo Cimento}, {\bf 11}, 467, (1974); J.D. Bekenstein [gr-qc/9710076]. 

\bibitem{sanvedholo} 
Sanved Kolekar, T. Padmanabhan,  \textit{Holography in Action}, [arXiv:1005.0619].
 
\end{thebibliography}
 \end{document}